\documentclass[11pt]{iopart}

\usepackage{iopams}
\usepackage{cases}
\usepackage{makeidx}
\makeindex
\usepackage {graphicx}
\def\>{\right\rangle}
\def\<{\left\langle}
\def\be{\begin{equation}}
\def\ee{\end{equation}}
\def\ba{\begin{array}{lll}}
\def\ea{\end{array}}
\def\f{\frac}
\def\dst{\displaystyle}
\def\beq{\begin{eqnarray}}
\def\eeq{\end{eqnarray}}
\begin{document}

\title{Neutral modes edge state dynamics through quantum point contacts
}
\author{\bf{D Ferraro}$^1$, A Braggio$^2$, N Magnoli$^1$ and M Sassetti$^2$}
\address{$^1$ Dipartimento di Fisica, Universit\`a di Genova, INFN, Via Dodecaneso 33, 16146 Genova, Italy}
\address{$^2$ Dipartimento di Fisica, Universit\`a di Genova, CNR--INFM LAMIA, Via Dodecaneso 33, 16146 Genova, Italy}
\begin{abstract}
  Dynamics of neutral modes for fractional quantum Hall states is
  investigated for a quantum point contact geometry in the weak-
  backscattering regime. The effective field theory introduced by Fradkin-Lopez
  for edge states in the Jain sequence is generalized to the case
  of propagating neutral modes. The dominant tunnelling processes are
  identified also in the presence of non-universal phenomena induced
  by interactions.  The crossover regime in the backscattering
  current between tunnelling of single-quasiparticles and of agglomerates of $p$-quasiparticles
  is analysed. We demonstrate that higher order cumulants of the backscattering
  current fluctuations are a unique
  resource to study quantitatively the competition between different carrier charges. 
  We find that propagating neutral modes are a necessary ingredient in order to explain this crossover phenomena.
\end{abstract}
\section{Introduction}
The Fractional Quantum Hall Effect (FQHE) represents one of the most
important examples of strongly correlated electron
systems~\cite{Laughlin83, Tsui82, DasSarma97}.  Due to the purely
two-dimensional electron motion a wider class of symmetries, for the exchange of identical particles, are
admissible in comparison with the three-dimensional
case~\cite{Myrheim99}.  This implies the existence of quantum fluids
with unique properties, such as fractional charges and
fractional statistics, and  eventually different internal topological
orders~\cite{Murthy03} at a given filling
factor~\cite{Stormer99}. In order to discriminate among these
different theories it is useful to investigate and characterize the edge
state dynamics~\cite{Moore09}.  Indeed, one may achieve unique
information about the bulk properties via the holographic
principle~\cite{Susskind95}, by studying the low energy gapless
excitations at the boundaries of the Hall bar~\cite{Wen90, Wen91}.

Edge states are described in terms of bosonic modes of a 
Chiral Luttinger Liquids ($\chi$LL)~\cite{Wen90b}.
For the description of the Laughlin sequence~\cite{Laughlin83}
with filling factor $\nu=1/(2m+1)$ ($m\in {\bf{N}}$), 
only a single charged mode is necessary. Here, theoretical predictions
of transport properties, as current and noise, through a Quantum Point
Contact (QPC) were proposed~\cite{Kane94, Chamon96} and confirmed by
experimental measurements~\cite{DePicciotto97, Saminadayar97}. It was
demonstrated that, in the weak-backscattering regime, the tunnelling
carriers have a fractional charge $e^*=\nu e$ ($e$ electron charge),
giving  the first confirmation of charge fractionalization~\cite{Laughlin83}. Further
experiments~\cite{Camino07} attempted to demonstrate also the anyonic
nature of these excitations~\cite{Wilczek90}.

For the Jain's sequence~\cite{Jain89} with $\nu=p/(2mp+1)$ and $p\in
{\bf{N}}$ one needs additional bosonic neutral modes in order to
fulfil the hierarchical structure~\cite{MacDonald90}.  As suggested
by Wen and Zee~\cite{Wen92}, one can introduce $p-1$ neutral
modes in addition to the charge one. The model predicts the expected value $e^* = \nu e /p$
for the charge of the fundamental quasi-particle (qp) excitation,
however, due to the increasing
number of fields, for increasing hierarchy $p$, the conserved currents
grows indefinitely \cite{Kane97, Cappelli02}, creating some problem of
consistency ~\cite{Lopez99}.  Some predictions of this theory
have been experimentally verified by investigating the transport
properties in cleaved-edge overgrown devices but, at the same time,
experiments shed light on the limits of the
model~\cite{Grayson98, Grayson06}.  In particular the behaviour of the
tunnelling current suggests that the dynamics
of neutral modes is less relevant with respect to the
prediction~\cite{Kane94, Kane95}. These problems triggered additional
theoretical investigations devoted to clarify the reasons of the
discrepancies ~\cite{Chang03}. Among them, it is important to mention
the model proposed by Fradkin-Lopez (FL model)~\cite{Lopez99}, which introduces only two neutral
topological modes (one for  infinite edge length~\cite{Chamon07}), 
instead of the above dynamical $p-1$ modes.  These
modes have zero velocity, then they do not contribute to the
dynamical response, having only a role in enforcing the appropriate
statistical properties.  
This model predicts also the correct value of the single-qp charge 
in agreement with the first experimental observations on tunnelling
through a QPC~\cite{Reznikov99}.  However, successive
experiments~\cite{Chung03, Heiblum03} have shown unexpected behaviour
in the conductance and, more intriguingly, a value $\sim pe^*$ for the carrier charge at extremely low
temperatures, larger then
the single-qp. These observations are not explainable within the framework of the original FL model.
Recently we suggested that these anomalies can be understood by
generalizing the FL model assuming propagating neutral
modes~\cite{Ferraro08}.  This explain the peculiar behaviour of the
experiments and the crossover regime between two different
carrier contributions: the single-qp with charge $e^*$ and the
$p$-agglomerate of qp with charge $p e^*$.

The role of neutral mode dynamics has been also addressed for other
FQHE states~\cite{Fiete07, Das09, Levkivskyi08} with emphasis on the concept of
finite velocity propagation~\cite{Levkivskyi08, Levkivskyi09, Wan08}.
It is believed that Coulomb interactions strongly increase the charge
mode velocity in comparison to the neutral one such that the two
modes energy cut-off bandwidths differ
consistently~\cite{Levkivskyi08, Levkivskyi09}. Recently, several
theoretical proposals on the possibility to detect the neutral mode 
velocity have been formulated. They range 
from the thermal Hall conductance~\cite{Kane97,
  Cappelli02} and the resonances in tunnelling through an extended
QPC~\cite{Overbosch09}, to the analysis of
Coulomb blockade peaks~\cite{Ilan08, Cappelli09, Cappelli09b}.  
Unfortunately, from experimental side the evidence of neutral modes 
is often missing or weak and quite indirect.\\

The work of the present paper is motivated by these open questions,
with the main task of giving evidence of the dynamics of propagating
neutral modes in transport properties.  In order to produce results
that can be tested experimentally we choose the standard geometry of a
QPC for the Jain series accessible to nowadays experimental
capabilities.  A possible renormalization of the dynamical response,
due to external interactions~\cite{Rosenow02, Yang03, Jolad09}, is also
taken in account, we will show that this introduces non-universal
behaviours.

In addition to current, we will analyze also current
fluctuations~\cite{Nazarov03} in order to properly describe tunnelling
of different charges.  The available theory of current cumulants for a
general tunnelling Hamiltonian~\cite{Levitov04}, combined with the
recent successes in measurements of current cumulants in
QPC~\cite{Reulet03,Bomze05} and other structures \cite{Gustavsson06,Fujisawa06}, gives a firm
background and a strong motivation for this analysis.  Current
cumulants~\cite{Levitov93,Levitov96,Braggio06} contain indeed several
information on the carrier charges, and may discriminate between
models based either on a single carrier or on many carriers,
considering a direct comparison with the available experimental data
on QPC in Hall bars~\cite{Chung03, Griffiths00, Rodriguez02, Heiblum06}.

We first find that neutral mode dynamics may be detected in the current
behaviour of single-qp for low enough temperatures.
We also demonstrate that weak interactions are
more favourable in order to detect the effects.

We point out that the most relevant effect, signal of the presence of
neutral modes, is the possibility to have different tunnelling
excitations by varying temperatures and voltages.  We prove the
existence of a crossover regime which separates the tunnelling
processes of $p$-agglomerates, at energies lower than
the neutral modes bandwidth, from the standard tunnelling of
single-qps at higher energies. We demonstrate that it is possible to find
voltages and temperatures regimes where the $p$-agglomerates are
detectable, although in presence of interactions and larger
temperatures these effects cannot be easily  visible.
For this reason we analyse noise and higher cumulants of the current fluctuations, whose behaviours 
can give clear information on the crossover regime.

In the shot noise regime the Fano factor is a proper quantity to detect the \emph{effective}
charge of the dominant carrier involved in tunnelling. We observe, as a
function of the external voltage, a double plateaux structure clearly
indicating the contribution of both the $p$-agglomerates and the
single-qp excitations.  In order to avoid the thermal regime, present
in noise measurements, we consider higher order odd cumulants, such as the
skewness.  These, indeed, are less sensible to temperatures and show
plateaux structures that can be interpreted in terms of dominant
contributions of different tunnelling charges. We show that the
comparison between different current cumulants behaviour can be a
unique resource to correctly interpret the physics of this crossover
phenomena~\cite{Chung03, Rodriguez02, Heiblum06, Dolev08}.

\noindent The outline of the paper is the following.  In
section~\ref{model} the generalization of the FL model to the case of
finite bandwidth neutral modes is presented and the field operator
construction of the edge excitations is derived with particular care
of the monodromy condition.  The scaling dimension is then discussed
in order to find the dominant tunnelling processes through the QPC.
In section~\ref{CurrentandNoise} we discuss analytical formulas for
transport properties via Keldysh technique at lowest order in
tunnelling.  The relation between current cumulants and backscattering
current is presented.  Results are discussed in
Sec.~\ref{result0} both on the backscattering current and on noise and
higher current cumulants.  Conclusions are presented in
Sec.~\ref{conclusion}. The~\ref{App:BetaM} contains the
detailed derivation of the field operator construction of the edge
excitations.
 
\section{Model}
\label{model}

We start by recalling the description of the edge states dynamics at the
boundaries of an Hall bar in the Jain sequence
\cite{Jain89} with filling factor $\nu=p/(2p+1)$ ($p\in\mathbb{N}$).
Within the minimal FL model ~\cite{Lopez99, Chamon07, Ferraro08}, edge states 
consist of two counter-propagating bosonic fields: a charged mode $\varphi^{c}(x)$ and 
a neutral one $\varphi^{n}(x)$.
 The corresponding Euclidean action for a given edges side is~\cite{Lopez99,Ferraro08} (from now on $\hbar=1$)
\begin{eqnarray}
\mathcal{S}^0&=&\f{1}{4\pi\nu}
\int\limits_{0}^{\beta} d\tau
\int_{-L/2}^{L/2}dx\,\, 
\partial_{x}\varphi^{c}(x,\tau)\left(i\partial_{\tau}+v_{c}\partial_{x}\right)\varphi^{c}(x,\tau)+
\nonumber\\
&+&\f{1}{4\pi}
\int\limits_{0}^{\beta} d\tau
\int_{-L/2}^{L/2}dx\,\, 
\partial_{x}\varphi^{n}(x,\tau)\left(-i\partial_{\tau}+v_{n}\partial_{x}\right)\varphi^{n}(x,\tau)\,,
\label{action}
\end{eqnarray}
where $\beta=(k_B T)^{-1}$ is the inverse temperature, and $L$ is the
edge length. We will consider the case $L\to \infty$
neglecting correction induced by the finite size.
Following~\cite{Ferraro08} the modes propagate with
different, but {\em finite} velocities: $v_{n}$ for the neutral mode
and $v_{c}$ for the charged one. The velocity of the charge mode can
be strongly increased by Coulomb interactions so we can safely
assume $v_{n}\ll v_{c}$~\cite{Levkivskyi08}.

As we have already shown in~\cite{Ferraro08}, neutral modes with
finite velocity affect qualitatively the transport dynamics and could
explain the observations of recent experiments~\cite{Chung03}.  We
will investigate this general case focusing on the main effects 
induced on transport properties.  Notice that neutral modes with
finite velocity differ from those described in the original FL model,
where they are \emph{topological} ($v_{n}=0$), therefore we denote the present 
model as Generalized Fradkin-Lopez model.

As clearly emerge from the action (\ref{action}) the two modes at the
same edge propagate in opposite directions.  Notice that, differently
from what happens in the hierarchical model \cite{Wen92}, where one
needs $p$ fields to describe each edge, in this minimal model the
number of fields is always two, independently on the value of filling
factor of the Jain series~\cite{Lopez99,Chamon07}.

The bosonic fields in (\ref{action}) can be represented in wave-number components as~\cite{Geller97, Merlo07}
\begin{equation}\label{phi}
  \varphi^{l}(x)=\sum_{k>0}\sqrt{\frac{2 \pi \nu_{l}}{k L}}
\left[a^{l}(k)^{}e^{i kx}+a^l(k)^\dagger e^{-i k x}\right] e^{-k a/2},
\end{equation}
where $k=2\pi s/L$ ($s \in\mathbb{N}$) are the wave vectors, $l=c,n$
denotes the charged and neutral modes and $\nu_c=\nu$, $\nu_n=1$. The
parameter $a^{-1}$ represents the ultraviolet wave-number cut-off that,
together with the mode velocities, determines the energy  
bandwidths $\omega_{c}=v_{c}/a$, $\omega_{n}=v_{n}/a$.
In the following, $\omega_{c}$ will be considered as the greatest energy scale 
with $\omega_n\ll\omega_c$.  The bosonic creation and annihilation
operators obey standard commutation relations
$[a^l(k)^{},a^{l'}(k')^\dagger] =\delta_{l,l'}\delta_{kk'}$.  They
ensure that the fields satisfy~\cite{Wen92, Lopez99} (for $L\to \infty$, $a\to 0$)
\begin{equation}
[\varphi^{c}(x),\varphi^{c}(y)]=i\pi
\nu{\rm sgn}(x-y)\,;\qquad [\varphi^{n}(x),\varphi^{n}(y)]=-i\pi{\rm sgn}(x-y).
\label{commutation}
\end{equation}

The electron number density along the edge is 
\be
\rho(x)=\f{1}{2\pi}\partial_{x}\varphi^{c}(x)
\ee
and depends only on the charge mode.

\subsection{Quasiparticle operators}
\label{Quasiparticle operators}
To complete the description one needs an
expression for the operators creating excitations at the edges. To be
as much general as possible we will discuss both the single-qp
excitations with charge $e^{*}=e\,(\nu/p)$ ($e$ the charge of the
electron) and multiple-qp excitations with charge $me^{*}$ with
$m>0$.\footnote{The operators with $m<0$ can be easily obtained from
  the operators with $m>0$ by exploiting particle-hole
  conjugation. For $m=0$ we have neutral excitations that do not impact on the current properties.}  We
refer in the following to these as $m$-agglomerates \cite{Ferraro08}.

There are three main constraints that the $m$-agglomerate operator $\Psi^{(m)}(x)$ 
has to satisfy~\cite{Ferraro08, Froehlich91, Ferraro09}
\begin{enumerate}
\item Charge: multiple excitations must have a charge of
  $me^{*}$. This implies appropriate commutation relation with the
  electron density

\be
\left[\rho(x),\Psi^{(m)}(y)\right]=-m\left(\f{\nu}{p}\right)\delta(x-y)\Psi^{(m)}(y)\,.
\label{charge}
\ee
\item Statistics: excitations have to satisfy fractional statistics 
\be
\Psi^{(m)}(x)\Psi^{(m)}(y)=\Psi^{(m)}(y)\Psi^{(m)}(x)e^{-i\theta_{m}\mathrm{sgn}(x-y)}\,,
\label{statistics}
\ee 
with the statistical angle~\cite{Wilczek90, Lopez99}
\be \theta_{m}=\pi
m^2\left(\f{\nu}{p^2}-\f{1}{p}-1 \right)-2\pi k.
\label{theta}
\ee 
The multiplicity
of $k\in \mathbb{Z}$ takes into account the $2\pi$ periodicity of the
phase. Note that this expression, given in the FL model, is in
accordance with the standard hierarchical construction
\cite{Wen92}.
\item Monodromy: the phase acquired by any excitations in a loop
  around an electron must be a multiple of
  $2\pi$~\cite{Froehlich97,Ino98}.
\end{enumerate} 
Using the bosonization technique the operator associated to the
$m$-agglomerate can be expressed
as an exponential of linear combinations of the bosonic edge fields~\cite{Ferraro08,Ferraro09}
\be
\Psi^{(m)}(x)=\f{\mathcal{F}^{(m)}}{\sqrt{2\pi a}}
e^{i\left[\alpha_{m}\varphi^{c}(x)+\beta_{m}\varphi^{n}(x)\right]}
\label{eq:quasiparticle_operator}\,. \ee
Here, $\mathcal{F}^{(m)}$ represent the so called Klein factors which will be discussed later, and the coefficients $\alpha_{m}$ and $\beta_{m}$ are determined by
imposing the fulfilment of the above three constraints. A detailed
discussions of the possible form of these coefficients can be found
in~\cite{Ferraro09} for $\nu=2/5$, and more generally in Appendix A for
$1<p\leq6$. Hereafter we simply quote the main results.
The relation (\ref{charge}) fixes 
$\alpha_{m}$. Indeed, using the commutation rule (\ref{commutation}) one finds directly
\be
\alpha_{m}=\f{m}{p}\,.\label{alfa}\ee

The second point (ii) is fulfilled by imposing the relation on the
neutral part\footnote{The over all sign in the expression for
  $\beta_{m}$ is arbitrary, the physical results on scaling dimension are not affected by this choice.}

\be
\beta_{m}=\sqrt{m^2\left(1+\f{1}{p}\right)+2k}\ .
\label{beta0}
\ee As one can see $\beta_{m}$ still depends on the free parameter $k$
in (\ref{theta}), showing that, for a given $m$, there is a family of
different excitations each with the same internal statistics.  From the square root in (\ref{beta0}) it follows \be
k\geq
k_{min}=-\mathrm{Int}\left[\f{m^2}{2}\left(1+\f{1}{p}\right)\right]\,,
\label{kappa}
\ee where $\mathrm{Int}\left[x\right]$ represents the integer part of
$x$~\footnote{Note that
  we changed definition of the coefficients $\alpha_{m}$ and
  $\beta_{m}$ and the sign of $k$ with respect to \cite{Ferraro08}.}.

The last requirement (iii) reduces even more the possible values of $k$.
As shown in Appendix A, for $p\leq 6$, it is possible to find the following form
\be
k=\f{p(p+1)}{2}(q^2-s^2) + d(p+1)(q-s)\,,
\label{kappa1}
\ee where $q\in\mathbb{Z}$, and $s$ and $d$ are integers that identify
the $m$-agglomerate for a given $p$ 
\be
m=sp+d,\qquad {\rm with}\qquad\qquad 0\le d \le p-1,\,\;\; {\rm and}\; s\ge
0\,.
\label{m}
\ee
Inserting the expressions (\ref{kappa1}) and (\ref{m}) in~(\ref{beta0}) one obtains
\be
\beta_m(q)=\sqrt{p(p+1)}\left(q+\frac{d}{p}\right)\,,
\label{beta1}
\ee where, for sake of clarity, we explicit the dependence on $q$.  We can now write the final expression of the
field operator, for a given $q$, within the family of the
$m$-agglomerate \ \be
\Psi^{(m,q)}(x,t)=\f{\mathcal{F}^{(m,q)}}{\sqrt{2\pi a}}
e^{i\left[(s+\f{d}{p})\varphi^{c}(x,t)+
    \sqrt{p(p+1)}(q+\f{d}{p})\varphi^{n}(x,t)\right]}\ .
\label{bosonicrepr}
\ee We conclude by commenting on the role of the Klein factors
${\mathcal{F}^{(m,q)}}$.  They are ``ladder operators'' that change
the particle numbers of the $(m,q)$-agglomerate type on the
edge~\cite{Haldane81, Haldane81b}. They ensure the appropriate statistical
properties between excitations on the same $m$-family, with different
$q$~\cite{Ferraro09} or on different edges~\cite{Guyon02}.  In the
following we will omit them since, apart their straightforward
ladder operation they do not play any role in the sequential tunnelling
regime treated in this paper~\cite{Guyon02, Martin05}.

\subsection{Scaling dimension} \label{secscalingdim} To investigate
the relevance of different $m$-agglomerates it is useful to consider
the behaviour of the two-point correlation function at imaginary
time~\cite{DasSarma97} \be \mathcal{G}^{(m,q)}(0,\tau)=\langle
T_{\tau}\Psi^{(m,q)}(0,\tau){\Psi^{\dagger}}^{(m,q)}(0,0) \rangle\,,
\label{Green_zeroT}
\ee where $T_{\tau}$ is the imaginary time ordering operator and
$\langle...\rangle$ the thermal average with respect to the free
action (\ref{action}). To evaluate (\ref{Green_zeroT}) one can use the bosonic representation
of the excitation field (\ref{bosonicrepr}) ending up with to two
distinct averages on charged and neutral modes
\begin{eqnarray}
\hskip-0.2cm\mathcal{G}^{(m,q)}(0,\tau)&=&\f{1}{2\pi a}
\langle T_{\tau}e^{i\alpha_{m}\varphi^{c}(0,\tau)}e^{-i\alpha_{m}\varphi^{c}(0,0)}\rangle
\langle T_{\tau}e^{i\beta_{m}(q)\varphi^{n}(0,\tau)}e^{-i\beta_{m}(q)\varphi^{n}(0,0)}\rangle\nonumber\\
&\equiv&\f{1}{2\pi a} e^{\alpha_m^2\mathcal{D}_{c}(0,\tau)} e^{\beta^2_m(q)\mathcal{D}_{n}(0,\tau)}\,.
\label{eq:green1}
\end{eqnarray}
Using standard identities for boson operators the above averages are represented as standard 
bosonic propagators $(l=c,n)$
\be
\label{eq:T0Green}
\mathcal{D}_{l}(0,\tau)=\langle T_{\tau}\varphi^{l}(0,\tau)\varphi^{l}(0,0)\rangle
-\langle(\varphi^{l}(0,0))^2\rangle\,.
\label{phononprop1}
\ee Since the Euclidean action (\ref{action}) is quadratic,
$\mathcal{D}_{l}(0,\tau)$ may be easily computed. For the present
discussion we only need the $T=0$ limit. One
has~\cite{DasSarma97, Chamon95} \be \mathcal{D}_{l}(0,\tau)=
-\nu_l\ln\left(1+\omega_{l}|\tau|\right),
\label{phononprop2_unren}
\ee   
where it is evident the role of the bandwidths $\omega_l$ as cut-off energies. 
Substituting this expression in (\ref{eq:green1}) one obtains the 
zero-temperature correlation function
\be
\mathcal{G}^{(m,q)}(0,\tau)=\frac{1}{2\pi a}\left( \frac{1}{1+\omega_c|\tau|}\right)^{\nu \alpha^2_m}
\left(\frac{1}{1+\omega_n|\tau|}\right)^{\beta^2_m(q)}\,.
\label{eq:Green_zeroTmodes}
\ee 
We can now compare the scaling properties between
different $(m,q)$-agglomerate operators.

First we recall that the relevance of a given excitation is
obtained by considering the scaling dimension $\Delta_m(q)$ defined as
half of the exponent of the correlation
function (\ref{eq:Green_zeroTmodes}) in the long time limit, at $T=0$. From~(\ref{eq:Green_zeroTmodes}) one has,  for $\tau\to\infty\ $ ~\cite{DasSarma97, Kane95}, 
\be \mathcal{G}^{(m,q)}(0,\tau)\propto |\tau|
^{-2\Delta_{m}(q)}\,.
\label{scalingdim} \ee The most relevant operator
is then the one with the minimal scaling dimension, this corresponds to a
dominance at long times and consequently at low energies.

The scaling dimension of
$(m,q)$-agglomerates is easily derived from
(\ref{eq:Green_zeroTmodes}) and~(\ref{scalingdim})
\be
\Delta_{m}(q)=\f{1}{2}\left[\nu\alpha_{m}^2+\beta^{2}_{m}(q)\right]=\f{1}{2}\left[\nu\left(s+\f{d}{p}\right)^2
  +p(p+1)\left(q+\f{d}{p}\right)^2\right].
\label{delta1}
\ee 
It depends both on charge and neutral modes
contributions.  

We are now in the position to determine the
most relevant operator at low energies.
Let us start to consider $\Delta_{m}(q)$ for a given $m$-family 
($s$ and $d$ are fixed by $m$ according to~(\ref{m})). 
It is easy to see that the most relevant operator among different $q$'s
is given by $q=0$ for $d/p\leq 1/2$  and $q=-1$ for $d/p>1/2$. 
Concerning the $d$ values there are two possible classes: $d=0$ which
corresponds to $s$ multiple of $p$-agglomerates ($m=sp$) and $d\ge 1$. For the
first class ($d=0$) the most relevant operator
corresponds  to $s=1$ with $m=p$, denoted as $p$-agglomerate with 
\be 
\Delta^{\rm min}_{p}=\f{1}{2}\nu < \f{1}{2}\,.
\label{deltaagg}
\ee 
For the second class ($d\ge 1$) the minimum scaling is reached for
$d=1$ and $s=0$, namely the single-qp with 
\be
\Delta^{\rm min}_{1}=\f{1}{2}\left[\f{\nu}{p^2}+\left(1+\f{1}{p}\right)\right]>\f{1}{2}\,.
\label{deltaquasip}
\ee 
Comparing (\ref{deltaagg}) and (\ref{deltaquasip}) we conclude
that the $p$-agglomerate is {\em always the most dominant operator}
at least for the discussed case with $1<p\leq 6$. This result covers the standard scaling region with energies
much smaller than both the neutral and the charged bandwidth $E=|\tau|^{-1}\ll \omega_{n},\omega_c$. 

In view of future analysis of transport properties it is also
important to discuss the intermediate energy region with
$\omega_{n}\ll E\ll\omega_c$, despite due to the finite energy values the
standard scaling argument has to be taken with more care. In the intermediate regime $\omega_{n}\ll |\tau|^{-1} \ll\omega_c$, the neutral modes  in
(\ref{eq:Green_zeroTmodes}) are
saturated and only charged modes contribute to the dynamics. It is than
convenient to introduce an \emph{effective} scaling dimension 
\be \Delta_{m}^{\rm
  eff}=\f{\nu\alpha_{m}^2}{2}=\f{\nu m^2}{2 p^2}\,,
\label{deltabar}
\ee that depends only on the $m$-agglomerate charge. At these
intermediate energies we then expect the dominance of the single-qp with
$m=1$, indeed $\Delta_{1}^{\rm eff}<\Delta^{\rm eff}_{p}$ for any
$p>1$. Then, the presence of a finite cut-off $\omega_n$ discriminates among
two different regions with a crossover from energies (e.g. voltages or
temperatures) smaller than $\omega_n$, where the $p$-agglomerate 
dominates, to energies larger than $\omega_n$ where the single-qp could
become more relevant.  

Before to conclude we would like to discuss the
robustness of the above results in the presence of possible
interaction effects due to additional external degrees of freedom. It is well known the possibility to have renormalization of the
dynamical exponents induced by a coupling with
external dissipative baths~\cite{Weiss99}. Different mechanisms of
renormalization were proposed ranging from coupling with
additional phonon modes~\cite{Rosenow02}, interaction effects
\cite{Papa04, Mandal02}, edge reconstruction~\cite{Yang03, Jolad09, Palacios96}.  In order
to qualitatively take in account these effects, we follow
\cite{Rosenow02}, considering baths given by one dimensional
phonon-like modes, coupled to charged and neutral modes at each
edges. As shown in detail in \cite{Rosenow02}
this interaction renormalizes the imaginary time bosonic propagators  
(\ref{phononprop1}) as
\be
\mathcal{D}_{l}(0,\tau)= -\nu_l g_l\ln\left(1+\omega_{l}|\tau|\right),\qquad
\label{phononprop2}
\ee with $g_c$ and $g_n$ interaction parameters. Note that $g_c$
correspond exactly to the factor $F$ defined
in~\cite{Rosenow02}. Here, we also assume renormalization
of the neutral modes described with the parameter $g_n$. Within this
model it is $g_{n,c}\ge 1$ with $g_{n,c}=1$ for the
unrenormalized case.  Note that the above renormalizations do not
affect the field statistical properties, which depend only on
the equal-time commutation relations, i.e. the field algebra. In our
discussion we simply consider the renormalization factors $g_c$ and $g_{n}$ as
free parameters.

Following analogous calculations, as done before, one can easily shown
that, at low energies, the $p$-agglomerate is still the most dominant
operator under the condition \be\label{interaction1}
\f{g_{n}}{g_{c}}>\nu\left(1-\f{1}{p}\right), \ee otherwise the
single-qp will dominate. Note that, from~(\ref{interaction1}), 
strong charge renormalizations may destroy the dominance of the
$p$-agglomerates in favor of the single-qp. However, the necessary
values of renormalizations must be so high, especially if some neutral
mode renormalization is also present, that this probably would not easily
happen in real cases. This explain why we expect that the $p$-agglomerate 
should be the most probable tunnelling entity at low energies in experiments. 
On the other hand, at intermediate energies
$\omega_{n}\ll E\ll\omega_c$, the relevance of the single-qp it is not
affected by the presence of interactions and it is then more robust.

\subsection{Tunnelling Hamiltonian}
Since the first measurements of transport properties on Hall bars
through QPC~\cite{DePicciotto97, Saminadayar97, Reznikov99}, 
tunnelling experiments revealed to be a precious tool to
investigate edge dynamics and internal properties of excitations.  In
this paper we consider a single QPC with
tunnelling between right and the left side edges. 

In the previous analysis we discussed only one edge,
now we have to extend it to the case of two different edges. We can
simply set the curvilinear abscissas on the edges such that 
right ($j=R$) and left ($j=L$) edge fields $\varphi_j^{l}(x)$ have
exactly the same field algebra as presented before for the single
edge.

Tunnelling of an $m$-agglomerate is described by the standard
Hamiltonian \be
\label{eq:tunnelling}
H^{(m)}_{T}=t_{m}\Psi^{(m)}_{R}(0){\Psi_L^{(m)}}^{\dagger}(0)+{\rm
  h.c.}, \ee where $\Psi^{(m)}_j(x_0)$ indicates the annihilation
operator of the $m$-agglomerate on the edge $j=R,L$ at point
$x_0$. This operator is defined in~(\ref{bosonicrepr})
for a generic edge. The coefficients $\alpha_m$ and $\beta_m$ 
will be chosen to have the minimal scaling dimension (\ref{delta1})
which, as we discussed before, corresponds to the most relevant
operator within a given $m$-family.  Without loss of generality we fix
the point of tunnelling at $x_0=0$.
The factors $t_{m}$ are the tunnelling amplitudes, they depend on the
detailed structure of the edges, not known in the framework of
effective theories, and on the precise geometry of the QPC, difficult
to be experimentally controlled. In the following we will define them
in terms of the single-qp tunnelling matrix element $t_1=\mathsf{t}$
such that $|t_{m}|=\kappa_m |\mathsf{t}|$, where the real factors
$\kappa_m$, represent the relative weight of the $m$-agglomerate
amplitude in comparison to the single-qp one ($\kappa_1=1$). These
parameters may be in principle fixed by fitting the experimental data.  

In the following sections we will calculate transport properties at
lowest order in the tunnelling, assuming the amplitudes $t_m$
sufficiently small to fulfil the weak-backscattering condition for
all possible $m$-agglomerates. In experiments one can appropriately
tune the QPC with pinch-off voltage to always satisfy the above
condition.  Following the previous discussion on the most relevant
operators in the different energy regime we will consider as
tunnelling processes only the two dominant contributions: the
single-qp and the $p$-agglomerate respectively.  The total tunnelling
Hamiltonian is then 
\be H_T=\sum_{m=1,p}H_T^{(m)}=\sum_{m=1,p}
t_m \ 
\Psi^{(m)}_{R}(0){\Psi_L^{(m)}}^{\dagger}(0)+{\rm h.c.}
\,.
\label{totaltunn}
\ee
As already discussed in~\cite{Ferraro08} this seems to be enough in order to 
explain the recent  experiments in the Jain series through QPC~\cite{Chung03}.

\section{Backscattering current and noise in QPC}
\label{CurrentandNoise}
\subsection{General Relations}
In the following, we derive general expressions for the tunnelling current and noise properties
of the generic $m$-agglomerate. We will make use of the Keldysh
formalism \cite{Martin05, Keldysh64, Kamenev09} without entering into the details and referring
to the literature for further discussions. At lowest
order in the amplitude, the possible tunnelling
processes described by (\ref{totaltunn}) are independent, and can be
evaluated separately. The backscattering current operator for an $m$-agglomerate is 
\be I^{(m)}(t)=i me^{*} \left[ t_{m} \,\,e^{i m e^{*} V
    t}\,\,{\Psi_R^{(m)}}^{\dagger}(t)\Psi_L^{(m)}(t)-{\rm h.c.}\right],
\label{current}
\ee where we adopted the interaction picture with the tunnelling Hamiltonian
(\ref{eq:tunnelling}) as interaction term.  Here, for convenience,
we introduced the effect of the bias $V$, applied to the QPC between
the two edges, in the amplitudes $t_m$ by the appropriate gauge
transformation $t_{m}\rightarrow t_{m} e^{im e^{*} V
  t}$~\cite{Martin05}.

The averaged backscattering current $\<\<I^{(m)}\>\>_1\!(t)$ can be
written in the Keldysh formalism as \be
\<\<I^{(m)}\>\>_1\!(t)=\f{1}{2}\sum_{\eta=\pm}\!
\<T_{K}\left[I^{(m)}(t^{\eta}) e^{-i\int_{K} dt'
    H^{(m)}_{T}(t')}\right]\>\,. \ee Here, the index $\eta=+,-$ label
the times on the Keldysh contour with $\eta=+$ for the forward branch
($-\infty, t$) and $\eta=-$ for the backward one ($t,-\infty$), the
exponential term contains the integral of the tunnelling Hamiltonian
in the interaction representation on the Keldysh contour $\int_K
dt$. The average is with respect to the thermal density matrix of the
unperturbed Hamiltonian at $t=-\infty$ and $T_{K}[...]$ is the Keldysh
time-ordering operator.

Expanding the exponential term one obtains the perturbative series of
the current in terms of the tunnelling amplitude. Inserting the
bosonized form~(\ref{bosonicrepr}) of the $m$-agglomerate operator and
performing standard thermal averages as discussed in (\ref{eq:green1}), one expresses the
steady ($t\to\infty$) backscattering current at lowest order
~\cite{Martin05} \be
\label{eq:current}
\!\!\<\<I^{(m) }\>\>_1\!=\dst-\f{i m e^{*} |\mathsf{t}|^{2}\kappa_m^2}{4\pi^2 a^2} 
\sum_{\eta=\pm}\!\eta\!\int^{+\infty}_{-\infty}\!\!\!\!\!\!\!dt' 
\sin(me^{*} Vt')\,\,e^{2\alpha^{2}_{m}{\mathcal{D}}^{\eta, -\eta}_{c}(t')}e^{2\beta^{2}_{m}
{\mathcal{D}}^{\eta,-\eta}_{n}(t')}\!,\\
\label{backscattering_current}
\ee where, ${\mathcal{D}}^{\eta,\eta'}_{l}(t)$ are the Keldysh Green's
functions relative to charged $l=c$ and to neutral $l=n$ modes.  They
are related to the standard greater Green's function
${\mathcal{D}}^>_l(t)=\<\varphi^{l}(t)\varphi^{l}(0)\>-\<(\varphi^{l}(0))^2\>$
in the following way \be {\mathcal{D}}^{\eta,\eta'}_{l}(t)= \cases{
  {\mathcal{D}}^>_l(\eta |t|) &
  for $\eta=\eta'$\\
  {\cal{D}}^>_l(\eta't) &
  for $\eta=-\eta'$\,.\\
} \label{Greenfun}\ee 
Note that the right and left edges give an equal form to
(\ref{Greenfun}).  This is taken into account by the factor 2 in the exponents
of~(\ref{backscattering_current}).  We recall that the greater Green's
function are directly related to the temperature correlators
introduced in~(\ref{eq:T0Green}) applying an analytic continuation to
real time~\cite{Chamon95,Mahan90}.

Let us now consider the backscattering current noise correlator defined as
\begin{equation}
\<\<I^{(m)}\>\>_2\!(t,t')=\f{1}{2}
\<\delta I^{(m)}(t)\delta I^{(m)}(t')+\delta I^{(m)}(t')\delta I^{(m)}(t)\>\,,
\end{equation}
with  $\delta I^{(m)}(t)\equiv I^{(m)}(t)-\<\<I^{(m)}\>\>_1(t)$ the current fluctuation.
In the Keldysh framework one has~\cite{Martin05}
\begin{eqnarray}
\<\<I^{(m)}\>\>_2\!(t,t')\!=\!
\f{1}{2} \sum_{\eta=\pm}\!
\<\!T_{K}\!\left[I^{(m)}(t^{\eta})I^{(m)}(t'^{-\eta})e^{-i\int_{K} dt' H^{(m)}_{T}(t')}\right]\!\>\nonumber\\
-\<\<I^{(m)}\>\>_{1}\!(t)\<\<I^{(m)}\>\>_{1}\!(t')\,.
\label{noise1}
\end{eqnarray}
The second order in $\mathsf{t}$ can be now
evaluated using the same procedure described for the current. In the
stationary limit the current noise correlator
depends on time's difference only, and the corresponding Fourier transform at zero frequency 
is \cite{Martin05}
\be
\!\!\<\<I^{(m)}\>\>_2=\f{(me^{*})^{2}
  |\mathsf{t}|^{2}\kappa_m^2}{4\pi^2 a^{2}} \sum_{\eta=\pm}\!
\int^{+\infty}_{-\infty}\!\!\!\!\!\!\!dt' \cos (m e^{*} Vt') e^{2\alpha^{2}_{m}
  {\cal D}^{\eta, -\eta}_{c}(t')}\,\,e^{2\beta_{m}^{2}{\cal
    D}^{\eta,-\eta}_{n}(t')}.  
\label{eq:noise}    
\ee 
By exploiting the analitycal properties of the bosonic Green's function 
(\ref{Greenfun}) in the
complex-time plane~\cite{Weiss99} \be
\label{eq:periodicity}
{\mathcal{D}}_{l}^{\eta,-\eta}(t+i\eta\beta/2)=\left[{\mathcal{D}}_l^{\eta,-\eta}(t-i\eta
  \beta/2)\right]^{*},\, \ee 
one can rewrite the time integral contour
in (\ref{backscattering_current}) and (\ref{eq:noise}) obtaining a relation, valid at the lowest order ~\cite{Martin05, Chamon95},
between the steady current
(\ref{backscattering_current}) and the zero frequency noise
(\ref{eq:noise})
\begin{equation}
\<\<I^{(m)}\>\>_2=me^{{*}} \left|\<\<I^{(m)}\>\>_1\right|\coth\left[
  \f{\beta m e^{*} V}{2}\right].
\label{Noise_current}
\end{equation}

It is convenient now to discuss the connection with the standard
perturbative approach of transport properties in terms of
tunnelling rates $\Gamma^{(m)}$.  The latter describe the probability
per unit time to have tunnelling of an $m$-agglomerate between the two
edges and can be easily calculated within Fermi golden rule at
lowest order in $\mathsf{t}$~\cite{Ferraro08, Weiss99} 
\be
\label{eq:rate}
\Gamma^{(m)}(E_m)=\f{ |\mathsf{t}|^{2}\kappa_m^2}{4\pi^2
  a^2}\int^{+\infty}_{-\infty}dt' e^{- iE_mt'}e^{2\alpha^{2}_{m}
  {\mathcal{D}}^>_{c}(t')}\,\,e^{2\beta_{m}^{2}{\mathcal{D}}^>_{n}(t')}\,,
\ee where $E_m=m e^{*} V$. As already pointed out by Levitov and
Reznikov in~\cite{Levitov04} the tunnelling
transport is described by a bidirectional Poissonian process. Therefore, there are two
rates: the forward $\Gamma^{(m)}(E_m)$ in (\ref{eq:rate}) and the
backward $\Gamma^{(m)}(-E_m)$, that corresponds to opposite
tunnelling processes. They are related via the detailed balance
condition $\Gamma^{(m)}(-E)=e^{-\beta E}\Gamma^{(m)}(E)$ as imposed by
the symmetry (\ref{eq:periodicity}). From the knowledge of the rates characterising
the bidirectional Poissonian process, one can directly calculate all
zero-frequency current cumulants~\cite{Levitov04} \be
\label{eq:cumul}
\langle\langle I^{(m)}\rangle\rangle_k=\cases{
(m e^{{*}})^k \left(1-e^{-\beta m e^{*} V}\right) \Gamma^{(m)}(E_m) & $k$ odd \\ 
(me^{{*}})^{k} \left(1+e^{-\beta m e^{*} V}\right)\Gamma^{(m)}(E_m) &  $k$ even.\\}
\ee 

Note that $k=1$ corresponds to the current (\ref{eq:current}), while
$k=2$ gives the noise (\ref{eq:noise}). We can now discuss the linear
regime $\beta m e^{*} V\ll 1$, the linear conductance is given by
\be
\label{eq:conductance}
G^{(m)}\equiv\lim_{V\to
  0}\frac{d\<\<I^{(m)}\>\>_1}{dV}=(me^{{*}})^{2}\ \beta\
\Gamma^{(m)}(0)\, , \ee 
and  as expected
from the fluctuation-dissipation theorem the thermal noise is 
$\<\<I^{(m)}\>\>_2\approx 2 k_B T G^{(m)}$.
In the opposite regime $\beta
m e^{*} V\gg 1$ the noise has the shot behaviour~\cite{Schottky18} with 
$\<\<I^{(m)}\>\>_2\approx m e^{*} \<\<I^{(m)}\>\>_{1}$.

Relations similar to (\ref{Noise_current}) can be obtained between the zero-frequency current
cumulants and the stationary
current. Here, 
we simply quote the final result 
\be
\label{pariedispari}
\langle\langle I^{(m)}\rangle\rangle_k=\cases{( m e^{*})^{k-1} \<\<I^{(m)}\>\>_1 & $k$ odd\\
  ( m e^{*})^{k-1} \left|\<\<I^{(m)}\>\>_1\right|\coth\left[
    \f{\beta m e^{*} V}{2}\right] & $k$ even. \\} \ee Notice that
all the previous relations are also applicable in the presence
of renormalization induced by coupling with external baths, since 
they are essentially based on the general analytic properties
(\ref{eq:periodicity}) of the thermal Green's functions.

Until now, we presented the cumulants for tunnelling current
 of $m$-agglomerates (see (\ref{eq:tunnelling})). As already
pointed out at lowest order, tunnelling of different agglomerates are
independent, and can be evaluated separately.  Considering the
tunnelling Hamiltonian~(\ref{totaltunn}) with the two most dominant
terms $m=1,p$ the current cumulants will be the sum of
the contributions of the two processes \be \langle\langle
I_{tot}\rangle\rangle_k=\sum_{m=1,p} \langle\langle
I^{(m)}\rangle\rangle_k\,.
\label{totalcurrent}\ee 
This independence will not hold at higher order in
the tunnelling.

\section{Results}
\label{result0}
\subsection{Single quasiparticle current}
\label{Single quasiparticle current}
In this part we will analyse in detail the tunnelling current of
single-qps with particular attention to the role of neutral modes.  We
remind that the single-qp operator is identified by $m=1$,
$d=1$ and $q=0$ in (\ref{alfa}) and (\ref{beta1}) such that
\be
\label{eq:alphabeta1} 
\alpha^2_1=\frac{1}{p^2}\,;\qquad
\beta^2_1=\left(1+\frac{1}{p}\right).  
\ee 

The first ingredient necessary for the evaluation of the
backscattering current (\ref{backscattering_current}) is the rate~(\ref{eq:rate}) that depends on the real time
finite temperature Green's functions of the bosonic modes
$\mathcal{D}_{l}^>(t)$. These functions are well known in
literature~\cite{Weiss99} and can be evaluated by analytic continuation from zero-temperature
imaginary time Green's function in (\ref{phononprop2}). Here, we
simply quote the result also in the presence of interactions $g_n,g_c$
\be
\label{eq:finiteT}
{\mathcal{D}}_{l}^>(t)=g_{l}\nu_l\ln{\left[\f{|\mathbf{\Gamma}\left(1+T/\omega_{l}-iT t\right)|^2}
{\mathbf{\Gamma}^2\left(1+T/\omega_{l}\right)\left(1-i\omega_{l}t\right)}\right],}
\ee
with $\mathbf{\Gamma}(x)$ the Euler gamma function (from now on $k_B=1$). 
The evaluation of the rate will be done by performing a numerical
time integration of~(\ref{eq:rate}). At zero temperature the rate can be also calculated 
analytically and the current is
given by~\cite{Ferraro08}
\begin{eqnarray}
\label{eq:qpcurrent}
\<\<I^{(1)}\>\>_1&=&\f{e^{{*}} |\mathsf{t}|^{2}}{2 \pi a^{2}} 
\f{e^{-E_1/\omega_{c}}}{\mathbf{\Gamma}(2\nu g_c\alpha^{2}_{1}+2g_n\beta_{1}^{2})E_1} \left(
  \f{E_1}{\omega_{c}}
\right)^{2\nu g_c\alpha^{2}_{1}}
\left(\f{E_1}{\omega_{n}}
\right)^{2 g_n\beta_{1}^{2}}\nonumber\\
&&\times\;_1F_1\left[2g_n\beta_{1}^{2},2\nu g_c\alpha^{2}_{1}+2g_n\beta_{1}^{2},\f{E_1}{\omega_{c}}-\f{E_1}
{\omega_{n}}\right],
\end{eqnarray}
with $E_1=e^{*} V$ and$\ _1F_1[a,b,z]$ the Kummer confluent 
hypergeometric function. Note that in case of neutral topological modes, where $\omega_n=0$, the
bosonic neutral mode propagator of~(\ref{eq:finiteT}) 
becomes ${\mathcal{D}}^{>,{\rm top}}_{n}=0$. The corresponding current at $T=0$ is 
\be
\label{eq:qpcurrenttop}
\<\<I^{(1,top)}\>\>_1= \f{e^{{*}} |\mathsf{t}|^{2}}{2 \pi a^{2}}
\f{e^{-E_1/\omega_{c}}}{\mathbf{\Gamma}(2\nu g_c\alpha^{2}_{1})E_1}
\left( \f{E_1}{\omega_{c}} \right)^{2\nu g_c\alpha^{2}_{1}}, \ee 
which corresponds to the limit $\omega_n\to0$ of (\ref{eq:qpcurrent}), showing 
the standard contribution of a single mode \cite{Weiss99, Kane92}.

\begin{figure}
\includegraphics[width=\textwidth,clip=true]{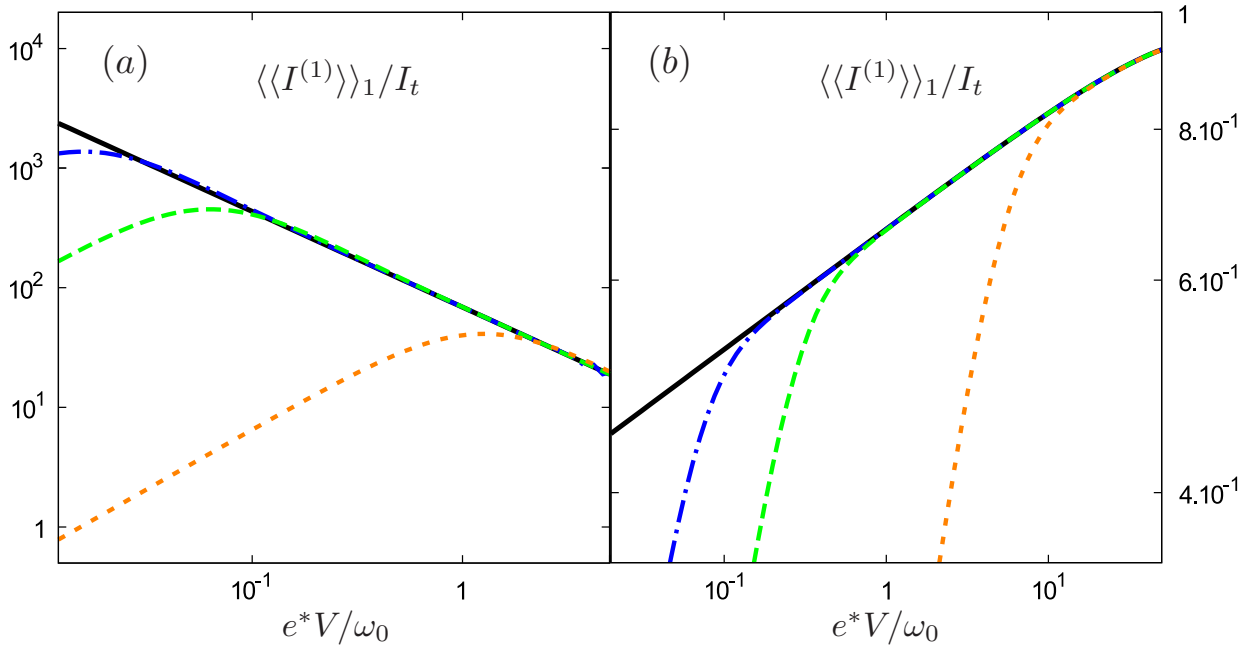}
\caption{Single-qp backscattering current as a function of the external
  voltage $e^{{*}}V/\omega_0$, with $\omega_0=10^{-3}\omega_{c}$, in bilogarithmic scale for $\nu=2/5$ and
  topological neutral modes $\omega_{n}= 0$.  (a) Unrenormalized case
  $g_c=1$; (b) renormalized case $g_{c}=5.5$. Different line styles
  correspond to temperatures: $T/\omega_0 =0$ (solid black), $0.025$
  (dot-dashed blue), $0.1$ (dashed green), $2$ (short dashed
  orange).
  The unit of the current is $I_t= e |\mathsf{t}|^{2}/(4\pi^{2} a^{2}
  \omega_c)$. The zero temperature power-laws of the figures are given
  by $V^{-4/5}$ and $V^{g_{c}/5-1}$ for the unrenormalized and the renormalized
  case respectively.}
\label{fig:IqpTop}
\end{figure}

\begin{figure}
\includegraphics[width=\textwidth,clip=true]{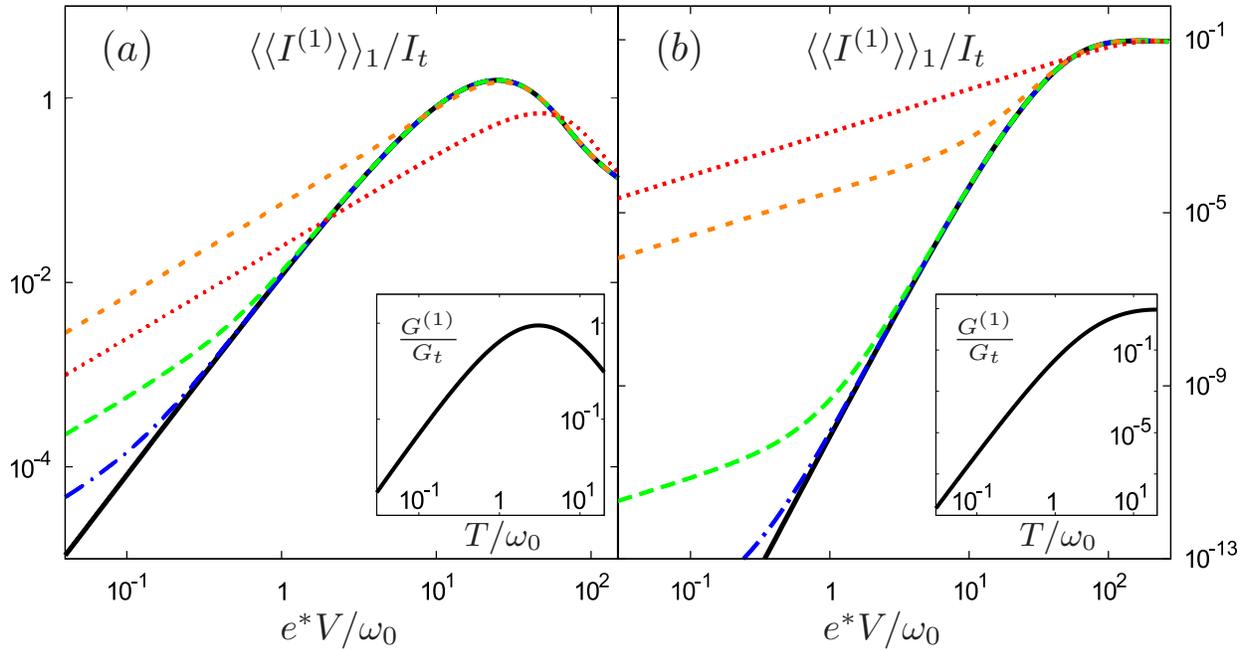}
\caption{Single-qp backscattering current as a function of the external
  voltage $e^{{*}}V/\omega_0$, with $\omega_0=10^{-3}\omega_{c}$, in bilogarithmic scale for $\nu=2/5$ and
  propagating neutral modes with $\omega_{n}=10^{-2}\omega_{c}$.  
(a) Unrenormalized case $g_c=g_n=1$; 
  (b) renormalized case $g_{c}=5.5$ and $g_n=2$. Different line styles correspond 
  to temperatures: $T/\omega_0 =0$ (solid black), 
$0.025$ (dot-dashed blue), $0.1$ (dashed green), 
$2$ (short dashed orange), $20$ (dotted red). 
At zero-temperature the power-laws scaling are:  
$V^{g_{c}/5+3 g_{n}-1}$ for $e^{*} V\ll \omega_n$ and  $V^{g_{c}/5-1}$ for $e^{*} V\gg \omega_n$.
The unit for the current is as in figure \ref{fig:IqpTop}.
In the inset the linear conductance in unit of $G_t=e^2
  |\mathsf{t}|^{2}/(4\pi^{2} a^{2}
  \omega_c^2)$ is represented in a bilogarithmic
  scale as a function of temperature.}
  \label{fig:Iqp}
\end{figure}

To discuss the effects of finite bandwidth, we compare the current
obtained for topological neutral modes in figure~\ref{fig:IqpTop},
with the one given in the presence of propagating neutral modes in
figure~\ref{fig:Iqp}.  We will analyse in detail the case of filling
factor $\nu=2/5$, with $\alpha_1^2=1/4$ and $\beta_1^2=3/2$.  However,
the results are qualitatively valid also for the other values of the Jain' sequence. We also consider  the presence
of renormalization.

Figure~\ref{fig:IqpTop} shows the voltage dependence of the
backscattering current for topological neutral mode ($\omega_n=0$) in
the unrenormalized case $g_c=1$ panel (a) and in the renormalized case
$g_c=5.5$ panel (b).  At zero temperature, black lines, one
recovers from~(\ref{eq:qpcurrenttop}) the power-law behaviour
 \be
\<\<I^{(1,top)}\>\>_1\propto V^{2\nu g_c \alpha_1^2-1},
\label{v1}
\ee driven by the charge exponent coefficient $\alpha_1$ and with a 
charge renormalization $g_c$.  For the unrenormalized case
figure~\ref{fig:IqpTop}(a) and for finite temperatures (coloured lines) the current exhibits a maximum around
$e^{*} V\approx T$ with a linear ohmic behaviour for smaller
voltages $e^{*} V \ll T$, and a power-law ~(\ref{v1}) at higher bias $e^{*} V \gg T$.

For sufficiently strong charge renormalization
$g_c>p^2/2\nu$ ($g_c>5$ for $\nu=2/5$), as shown in
~\ref{fig:IqpTop}(b), the power-law exponent in~(\ref{v1}) is positive and
the current is always growing. Then, for finite temperatures the maximum
is substituted by a smooth change of power-laws between the ohmic
regime and the zero temperature
behaviour~(\ref{v1}).

We can now compare the above behaviour with propagating neutral modes as shown in
figures~\ref{fig:Iqp}(a) and \ref{fig:Iqp}(b).  Let us start with the
zero temperature case, here the current displays two different scaling
\begin{numcases}
{\<\<I^{(1)}\>\>_1\propto}V^{2\nu g_c \alpha_1^2+2
    g_n
    \beta_1^2-1}
    & $e^{*} V\ll \omega_n$\label{vv01}\\
  V^{2\nu g_c \alpha_1^2-1}
  & $e^{*} V\gg \omega_n$.\label{vv02} 
\end{numcases}
Neutral modes contribute to the dynamics only for voltages
smaller than the bandwidth $\omega_n$. The power-law exponent 
(\ref{vv01}) is always positive because of (\ref{eq:alphabeta1}) and
$g_c,g_n\geq 1$.  Only at higher voltages $e^{*} V\gg \omega_n$, as in (\ref{vv02}), 
one recovers the standard FL
behaviour of the single-qp current where the dynamical
contribution of neutral modes is absent (cf. (\ref{v1})). For weak
renormalization\footnote{Note that the classification of weak and 
  strong renormalizations adopted here is limited to the analysis of  
  single-qp current. Later we will consider different definitions in order to 
follow the richer phenomenology introduced by the presence of the $p$-agglomerates.} 
$g_c<p^2/2\nu$ ($g_c<5$ for $\nu=2/5$), as represented
in figure~\ref{fig:Iqp}(a), a maximum at $V=V^{0}_{\rm max}$ separates
the two above regimes, while for strong renormalization $g_c>p^2/2\nu$ as in figure~\ref{fig:Iqp}(b), the maximum is replaced by a
smooth crossover at $V=V_{\rm c}^{0}$ with a growing current. The two
quantities $V^{0}_{\rm max}$ and $V_{\rm c}^{0}$ are defined at zero
temperature as their notation remind us.
In general both $V^{0}_{\rm max}$ and  $V_{\rm c}^{0}$ are proportional to $\omega_n$. 
However, their precise values depend on the parameters such as the
filling factor and the renormalizations. For example,
$V^{0}_{\rm max}$ 
can be obtained from~(\ref{eq:qpcurrent}), 
in the limit $\omega_{c}\to\infty$, by solving the equation \be
\label{eq:maxcurr}
\ _1F_1\left[2g_n\beta^{2}_{1},2\nu
  g_c\alpha^{2}_{1}+2g_n\beta^{2}_{1}-1,-e^{*}V^0_{\rm
    max}/\omega_{n}\right]=0\,.  \ee For $\nu=2/5$ and $g_n=g_c=1$, we
find $\e^{*} V^0_{\rm max}\approx2.5\,\omega_n$ in accordance with
the position of the maximum in figure~\ref{fig:Iqp}(a). 

At finite temperatures and low voltages $e^*V\ll T$ the current shows a linear
behaviour $\<\<I^{(1)}\>\>_1\approx G^{(1)}V$ proportional to the
linear conductance $G^{(1)}$. The latter is represented in the insets of the
figure~\ref{fig:Iqp}, at temperatures
smaller than the neutral bandwidth, $T\ll \omega_n$, neutral modes
affect its scaling with $G^{(1)}\propto T^{2(\nu g_c \alpha_1^2+ g_n
  \beta_1^2-1)}$. On the other hand, at $T\gg \omega_n$ the scaling is
driven by the charged mode only with $G^{(1)}\propto T^{2(\nu g_c
  \alpha_1^2-1)}$~\cite{Ferraro08}. These two power-laws are
distinguishable in the
curves presented in the insets.

At larger voltages the scaling behaviour~(\ref{vv01}) with neutral
modes is still visible for voltages range $T/e^{*} \ll V \ll
V^{0}_{\rm max}$ for weak renormalizations (figure~\ref{fig:Iqp}(a))
or for $T/e^{*} \ll V\ll V^{0}_{\rm c}$ for the strong case
(figure~\ref{fig:Iqp}(b)).  Under these conditions the maximum and the
crossover position remain substantially \emph{uneffected} varying
the temperature. Only at higher temperatures $T/e^{*}\gtrsim V^{0}_{\rm
  max}$ or $T/e^{*}\gtrsim V^{0}_{\rm c}$, depending on the renormalization,
the neutral modes are saturated at any voltages. Their dynamical
contributions disappear and the current directly pass from the linear
behaviour to the power-law~(\ref{vv02}) determined by the charged
modes only.  In these high temperatures condition, represented in
figure~\ref{fig:Iqp} with orange and red lines,
the position of the maximum $V_{\rm max}$ or the crossover $V_{\rm c}$ starts to
increase linearly with temperature.

The above discussion shows two important aspects:
\begin{itemize}
\item[i)] renormalization effects change qualitatively the current behaviour;
\item[ii)] neutral modes dynamics may be visible in the current.
\end{itemize}
In particular, we have seen that weak interactions may be most favourable
in order to achieve point ii). Indeed, the
experimental verification of a {\em temperature independent} maximum
position, which would be a trade-mark of the presence of neutral modes
is more easily detectable than the measurement of a temperature
independent crossover that would be the case for strong
renormalization.

We would like to conclude by commenting that if it seems not
too difficult to give evidence of the neutral modes, much more delicate
is the possibility to quantitative determine the neutral bandwidth
$\omega_n$, since the current behaviour depends crucially on other a priori unknown
parameters.

\subsection{Single-qps and $p$-agglomerates crossover phenomena}
In this part we will consider the behaviour of the total
backscattering current $\<\<I_{tot}\>\>_1$in ~(\ref{totalcurrent})
given by the current contributions of the two most relevant processes: the
single-qp $\<\<I^{(1)}\>\>_1$, and the $p$-agglomerate 
$\<\<I^{(p)}\>\>_1$. The aim will be to investigate the crossover
phenomena among these two different terms. We remind that the weight of
these tunnelling processes are given by $t_1\equiv\mathsf{t}$ and
$|t_{p}|=\kappa_p |\mathsf{t}|$. In the following $\kappa_p$ will be
assumed as a free parameter. Since the experimental data, such as \cite{Chung03}, 
shows the presence of single-qp at larger voltages/temperatures and the
presence of $p$-agglomerates only for lower energies one need to choose
$\kappa_p$ sufficiently low such that the effects of the
$p$-agglomerate are not visible at high energies but it has to be
big enough to see them at sufficiently low energies.  

In order to determine $\<\<I^{(p)}\>\>_1$ we remind that the most
relevant $p$-agglomerate operator in~(\ref{bosonicrepr}) is given by
$d=q=0$. This implies $\beta_{p}=0$ in~(\ref{beta1}), with {\em no
  presence of neutral modes} in the dynamics. Only charged modes will
enter with the coefficient $\alpha_{p}=1$ as stated in~(\ref{alfa}).
This allows to determine the current $\<\<I^{(p)}\>\>_1$ by using the
results obtained in the previous section for the single-qp process
with topological neutral modes $\omega_n=0$ -
see~(\ref{eq:qpcurrenttop}) - apart from the necessary substitution
$\alpha_{1}\to \alpha_p=1$.

\begin{figure}[h]
\includegraphics[width=\textwidth,clip=true]{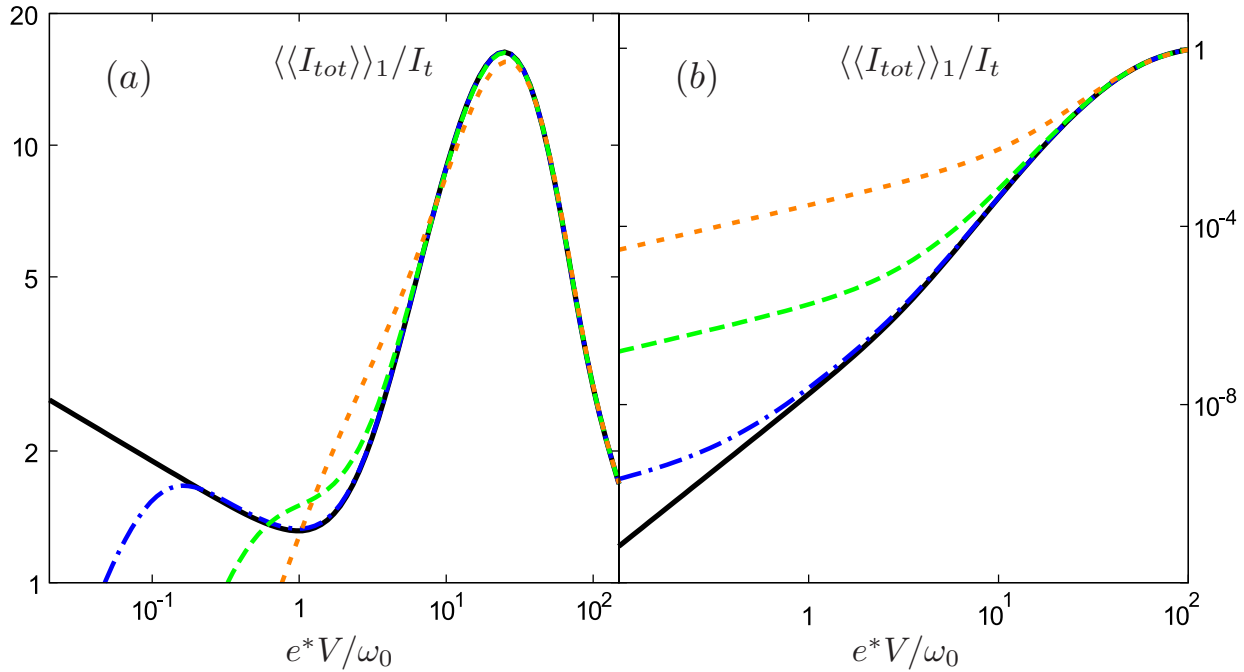}
\caption{Total backscattering current $\<\<I_{tot}\>\>_1$ as a
  function of $e^{{*}}V/\omega_0$, with $\omega_0=10^{-3}\omega_{c}$, in bilogarithmic scale for $\nu=2/5$. (a) 
  Unrenormalized case $g_c=g_n=1$ with $\kappa_p=0.4$; (b)
  renormalized case $g_{c}=5.5$ and $g_{n}=2$, with $\kappa_p=10$. Temperatures are
  $T/\omega_0 =0$ (solid black), 
$0.025$ (dot-dashed blue), $0.1$ (dashed green), $2$ (short dashed orange). 
The neutral mode bandwidth is $\omega_{n}=10^{-2}\omega_{c}$. 
Unit of current:  
$I_t= e |\mathsf{t}|^{2}/(4\pi^{2} a^{2}\omega_c)$.} 
\label{fig:ItotUnren}
\end{figure}
Figures~\ref{fig:ItotUnren}(a) and~\ref{fig:ItotUnren}(b) show
the total backscattering current as a
function of voltage for unrenormalized and
renormalized cases respectively, and $\nu=2/5$. 

We start the discussion at zero temperature corresponding to black lines. The current
shows three different power-laws 
\begin{numcases}{\<\<I_{tot}\>\>_1\propto}
V^{2\nu g_c\alpha_p^2-1}
& $V\ll V^*$\hskip2cm\label{tot1}\\
V^{2\nu g_c \alpha_1^2+2g_n \beta_1^2-1}
\,& 
$V^*\ll V\ll \omega_n/e^*$\label{tot2}\\
V^{2 \nu g_c \alpha_1^2-1}
& $\omega_n/e^*\ll V$,
\label{tot3}
\end{numcases}
with $\alpha_p$, $\alpha_1$ in~(\ref{alfa}) and $\beta_1$
in~(\ref{eq:alphabeta1}).  Here, we defined with $V^*$ the bias at
which the two main current contributions, at zero temperature, are
equal $\<\<I^{(p)}\>\>_1=\<\<I^{(1)}\>\>_1$.  The precise value of
this point crucially depends on $\kappa_p$ and on other
parameters.

 Note that 
the power-laws~(\ref{tot1})-(\ref{tot3}) depend on the parameters
$g_{c,n}$ which may affect qualitatively the behaviour. We
already discuss their influence on the single-qp contribution, here
we would like to focus mainly on the $p$-agglomerate part, which dominates the current at low voltages, see~(\ref{tot1}), and on the crossover 
region.

For weak values of charge renormalizations $1\leq g_c< 1/2\nu$
the current decreases with voltages at $V<V^*$, see~(\ref{tot1}) and
the black line in figure~\ref{fig:ItotUnren}(a) present a minumum around $V^{*}$. This behaviour
is completely different from the case of larger renormalizations
$g_c>1/2\nu$, which shows an increasing current
(cf. figure~\ref{fig:ItotUnren}(b)). In this last case, there is a
tendency to lose the clear evidence of the $p$-agglomerate which
merges very smoothly on the single-qp contribution.  Indeed, the
crossover is only signalled by a change in the power-law that is
gradually less evident increasing charge renormalization. Hence, we can conclude that, at $T=0$, the
weak interaction case with $1\leq g_c< 1/2\nu$, which exhibits a
minimum in the current, is the most favourable in order to show
the presence of $p$-agglomerates.

We would like now to investigate the robustness of the above results
for finite temperature. The main task is to extract the
voltage and temperature ranges where one
could clearly detect the contribution of $p$-agglomerates. At
finite temperatures (colored lines in
figure~\ref{fig:ItotUnren} ), and low bias $V\ll T/pe^{*} < V^*$ the
current, instead of the power-law~(\ref{tot1}), shows a 
linear behaviour typical of the ohmic regime
preventing divergence at very low voltages for weak interactions ($g_c<1/2\nu$). On the other hand at larger voltages $T/pe^*\ll V< V^*$ it is possible to detect the presence of $p$-agglomerates 
from the behaviour of the current that follows the power-law ~(\ref{tot1}).  At larger voltages $V\gg  V^*$, single-qp dominates and current is  similar to the $T=0$ cases 
(\ref{tot2}) and (\ref{tot3}). 

As a common trend we observe that the optimal possibility to detect
agglomerates depends crucially on the interaction parameters
$g_{c,n}$.  From figure~\ref{fig:ItotUnren}(a), representative of weak
renormalization $g_c<1/2\nu$, one can see that the transition between the
linear behaviour to the $p$-agglomerates scaling can create a peak
around $p e^{*} V\approx T$.  In this case the current exhibits two
peaks: the first due to the crossover between ohmic behaviour and the $p$-agglomerate power-law, the second due to the crossover inside the single-qp
contribution due to the de-activation of neutral modes. Increasing
interactions ( $g_c>1/2\nu$) the first peak tends to disappear until
for strong enough interactions, $g_c>p^2/2\nu$, as in figure
\ref{fig:ItotUnren}(b), even the second maximum disappears in
accordance with the behaviour of the single-qp contribution discussed
in the previous section. Here, the change in the power-law induced by the appearance/disappearance
of $p$-agglomerates becomes less visible.

The above discussions demonstrate that is possible to find voltages
and temperatures regimes where the agglomerate contributions are
detectable in the current. In order to confirm this important result we would like to find
stronger signatures in other transport
properties.  The study of higher
current moments, performed in the next section, will represent a
powerful tool to clarify the crossover physics between the
$p$-agglomerates and single-qps.

\subsection{Fano factor and higher cumulants}
Noise measurements are important tools in order to obtain clear
information on the carrier charges involved in transport and
consequently on the dominant excitations present in the tunnelling
current.

In order to study the current fluctuations we start by considering the Fano
factor $F_2=\<\<I_{tot}\>\>_2/e |\<\<I_{tot}\>\>_1|$ defined as the
ratio between the zero frequency noise of the total backscattering
current and the current itself.
Using~(\ref{Noise_current}) one can write the Fano in terms of single-qp and $p$-agglomerate current contributions
\begin{equation}
\hskip-0.5cm F_2=\frac{e^{*}}{e}\frac{\<\<I^{(1)}\>\>_1
\coth(\beta e^{*} V/2)+ p\<\<I^{(p)}\>\>_1
\coth(\beta pe^{*} V/2)}{|\<\<I^{(1)}\>\>_1+\<\<I^{(p)}\>\>_1|}\,.
\label{fano}
\end{equation} 
In the following we will restrict the discussion at energies smaller
than the neutral bandwidth, i.e. $e^{*} V,T\lesssim\omega_n$, for
which crossover phenomena appear.
\begin{figure}
\includegraphics[width=\textwidth,clip=true]{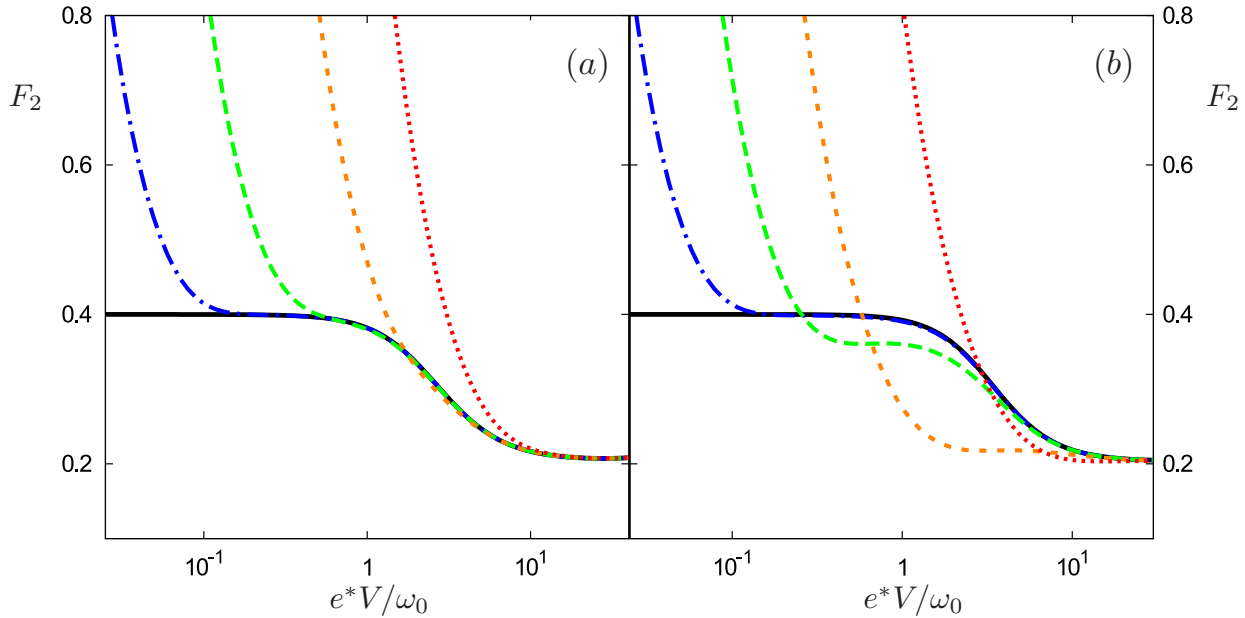}
\caption{Fano factor $F_2$ as a function of bias $e^{*} V/\omega_0$, with $\omega_0=10^{-3}\omega_{c}$, in linear-logarithmic  scale for $\nu=2/5$ and for 
$\omega_n=10^{-2}\omega_c$.  (a) Unrenormalized case
  $g_c=g_n=1$, with $\kappa_{p}=0.4$; (b) renormalized case
  $g_{c}=5.5$ and $g_{n}=2$, with $\kappa_{p}=10$. Different line styles
  correspond to temperatures:  $T/\omega_0 =0$ (solid black), 
$0.025$ (dot-dashed blue), $0.1$ (dashed green), $0.5$ (short dashed orange), $2$ (dotted red).}
\label{fig:Fano}
\end{figure}
 Figure~\ref{fig:Fano} shows the Fano factor as a function of bias at
 $\nu=2/5$ in the unrenormalized case, panel~\ref{fig:Fano}(a), and
 in the strong renormalized case, panel~\ref{fig:Fano}(b).  At zero
 temperature (black lines) the system is in the shot-noise
 regime, from~(\ref{fano}) one then has 
\begin{equation}
F^0_2=\frac{e^{*}}{e}\frac{\<\<I^{(1)}\>\>_1
+ p\<\<I^{(p)}\>\>_1}{\<\<I^{(1)}\>\>_1+\<\<I^{(p)}\>\>_1}\,.
\label{fano1}
\end{equation}
We remind that the total current in~(\ref{tot1})-(\ref{tot3})  is dominated by the 
$p$-agglomerates for $V\ll V^*$ where $\<\<I_{tot}\>\>_1\approx
\<\<I^{(p)}\>\>_1$, and by the single-qp at
$V\gg V^*$ where $\<\<I_{tot}\>\>_1\approx \<\<I^{(1)}\>\>_1$.
Hence, the Fano shows two different plateaux 
\begin{numcases}{F^{0}_2\approx}
\f{pe^{*}}{e}=\nu\, & $V\ll V^*$\hskip2cm\label{fanozero1}\\
\f{e^{*}}{e}=\f{\nu}{p}\, & $V\gg V^*$.\label{fanozero2}
\end{numcases}

At low bias - cf.~(\ref{fanozero1}) - the Fano is dominated by $p$-agglomerates 
with effective charge $p e^{*}$, while at higher
voltages - cf.~(\ref{fanozero2}) - the single-qp prevails with charge
$e^{*}$.  This behaviour is clearly showed in
figures~\ref{fig:Fano}(a) and~\ref{fig:Fano}(b) and is stable against
renormalization effects. Low and high bias regimes are
separated by an interpolating region. As shown in~\cite{Ferraro08} the
width of this region is not universal and crucially depends on the interaction strength.

Note that the presence of two plateaux in the Fano is a direct
evidence of the existence of two different kind of carriers that participate
to tunnelling. At the zero temperature their charges are deduced by
the values of the plateaux themselves.
 
In order to compare with experiments we need to generalize the above results at finite
temperatures. We will see that some previous conclusions are still valid
despite the presence of more intricate regimes.
At small voltages $p e^{*} V \ll T$ the noise is represented by the Johnson-Nyquist thermal
contribution with $F_2=2k_BT/eV$ and is diverging for $V\to 0$. This is clearly visible in all the
finite temperature curves of figure~\ref{fig:Fano}.
For the discussion at larger voltages we consider separately weak and strong renormalizations.  Let us start with
figure~\ref{fig:Fano}(a).  For temperatures smaller than
the threshold voltage $V^*$ and voltages in the range
$T/pe^{*} \ll V<V^*$ the Fano reaches the first higher plateau at
$F_2\approx \nu$ related to the charge of the $p$-agglomerate. 
On the other hand, at higher bias, it drops to the second plateau at
$F_2\approx\nu/p$ corresponding to the charge of the single-qp.  

At larger temperatures $T\gtrsim e^{*} V^*$ the visibility of the 
$p$-agglomerate is compromised by the presence of thermal noise. Indeed,
the Fano directly passes from the Johnson-Nyquist regime to the
single-qp plateau (see e.g. orange and red curves
in~figure~\ref{fig:Fano}(a)) without a possibility to detect the $p$-agglomerate
plateau. One then can say that, for weak interaction one may
detect the presence of two different excitations with a proper
determination of charges only if $T \ll e^{*} V^*$, that typically in 
the experiments correspond to very low temperatures.

For strong renormalizations, as in figure~\ref{fig:Fano}(b), the
results are even more involved.  While the extremely low and high
temperature regimes with $T\ll e^{*} V^*$ and $T\gg e^{*} V^*$ show
a similar behaviour as observed above, at intermediate temperatures
qualitatively differences appear.  For example green line in
figure~\ref{fig:Fano}(b) shows two plateaux, but the first is at an intermediate
value $\nu/p\leq F_2^*\leq\nu$ smaller than the $p$-agglomerate
charge.  The origin of this behaviour is due to the smooth crossover
between the $p$-agglomerate and the single-qp contributions present
for strong interaction.  Indeed, differently from the unrenormalized
case, there is a finite region of bias where the two current
contributions are comparable.  Hence, the plateau value at
$F_2^*$ corresponds to a weighted average of the single-qp
$(e^{*}/e)$ and $p$-agglomerate
$(pe^{*}/e)$.

We can then summarize the following important results:

\begin{itemize}
\item[i)] the existence of more than one plateau in the Fano factor
  is a clear trademark of the presence of more than one relevant tunnelling
  entity;

\item[ii)] whenever more than one relevant tunnelling process is
  simultaneously present {\em is not} appropriate to evaluate the
  effective charge involved in tunnelling by fitting the noise
  in~(\ref{Noise_current}) using a single carrier charge as free parameter.
\end{itemize}

This second procedure is often used in experiments
\cite{DePicciotto97, Saminadayar97, Reznikov99, Chung03}, however it
could bring to misinterpretations. For example, in the above case with
an intermediate plateau at $F_2^*$, the current is still composed by
two different excitations with different charges equally important.
In this case fitting the Fano with a single \emph{effective} charge
will give a wrong description.
All these observations could be relevant for the
experiment in~\cite{Chung03} especially for $\nu=3/7$, where
apparently the Fano factor instead to be equal to $3/7$, even at
low temperatures, reaches only the value of $2.4/7$.  

\begin{figure}
\includegraphics[width=\textwidth,clip=true]{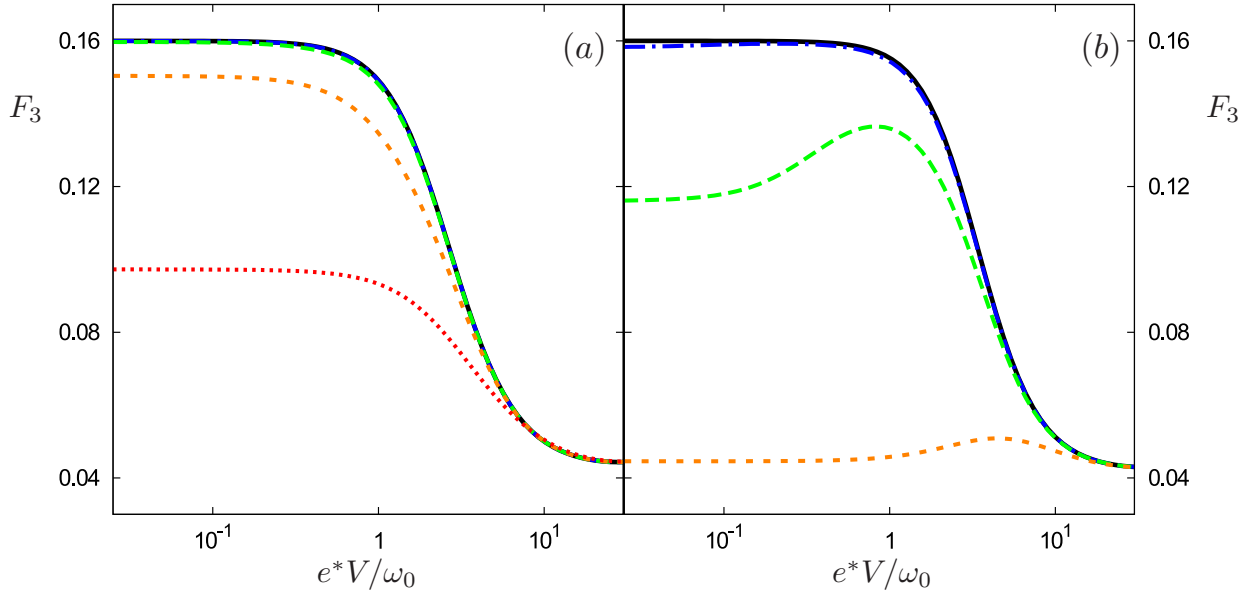}
\caption{Normalized skewness $F_3$ as a function of bias $e^{*} V/\omega_{0}$, with $\omega_0=10^{-3}\omega_{c}$,
  in linear-logarithmic scale for $\nu=2/5$ and 
  $\omega_n=10^{-2}\omega_c$.  (a) Unrenormalized case $g_c=g_n=1$
  with $\kappa_{p}=0.4$; (b) renormalized case $g_{c}=5.5$ and
  $g_{n}=2$ with $\kappa_{p}=10$.  The different line styles
  correspond to different temperatures according the same convention
  adopted in figure~\ref{fig:Fano}. The units are the same as figure~\ref{fig:Fano}.}
\label{fig:Skewness}
\end{figure}

In this last part of the section we would like to generalize the above discussion
considering higher current cumulants. In view of their statistical 
independence, we expect to obtain complementary information on carrier charges from them.
We define the generalized Fano factors as the normalized $k$-th
order zero frequency current cumulants~\cite{Braggio06}
\be
\label{oddcum}
F_k=\f{\<\<I_{tot}\>\>_k}{e^{k-1} \<\<I_{tot}\>\>_1}, 
\ee 
with $\<\<I_{tot}\>\>_k$ the total cumulant defined in~(\ref{totalcurrent}).
In the following we will consider the behaviour of odd cumulants with
a detailed analysis of the normalized skewness $F_3$. This quantity is
indeed experimentally accessible for a QPC in a Hall
bar~\cite{Reulet03,Bomze05}.  The general expression for odd $k$-th normalized cumulants can be
easily obtained inserting~(\ref{pariedispari}) and~(\ref{totalcurrent}) in~(\ref{oddcum})
\begin{equation}
\label{eq:Fksharing}
F_k=\left(\frac{e^{*}}{e}\right)^{k-1}
\f{\<\<I^{(1)}\>\>_1+(p)^{k-1}\<\<I^{(p)}\>\>_1}{
  \<\<I_{tot}\>\>_1}\ .
\end{equation} 
Note that this expression also holds at finite temperatures for odd
$k$-th cumulant. For even $k$-th cumulants a $\coth(m e^* V/T)$ term 
appears such
that (\ref{eq:Fksharing}) is only valid at low
temperatures $T\ll e^{*} V$.
The absence of this term in the odd cumulants avoids
divergences induced by the thermal regime which are, on the other hand,
always present for even cumulants.  Following similar analysis as done
for the Fano we can extract the zero
temperature behaviour of~(\ref{eq:Fksharing})
\begin{numcases}{F^{(0)}_k\approx}
\left(\f{pe^*}{e}\right)^{k-1}=\nu\,^{k-1}\, & $V\ll V^*$\hskip2cm\label{zerofano1}\\
\left(\f{e^*}{e}\right)^{k-1}=\left(\f{\nu}{p}\right)^{k-1}\, & $V\gg V^*$\,.\label{zerofano2}
\end{numcases}
Again two plateaux appear with values related to the $p$-agglomerate charge at low
voltages and to the single-qp charge at high voltages, with the power $(k-1)$. The skewness in
figure~\ref{fig:Skewness} reproduce this behaviour (black lines) both for the unrenormalized case of figure~\ref{fig:Skewness}(a) and for
the strong renormalization of figure~\ref{fig:Skewness}(b).

Increasing temperature the two plateaux are still present with a
decrease of the higher with respect to the $T=0$ value.  
Note that the thermal
divergence is not present here.
Therefore, from the odd cumulants behaviour,  one can have information on different
excitations (looking at the different plateaux) even in the deep
thermal regime $e^{*} V\ll T$ which is not an observable regime in noise measurements.

To be more specific let us compare the two red curves in
figures~\ref{fig:Skewness}(a) and~\ref{fig:Fano}(a).  While for the Fano the thermal regime cancels
out completely the presence of the higher plateau, this one is still
clearly visible in the skewness. This means that odd moments can be an
important tool in order to detect the presence of different
excitations (e.g. the $p$-agglomerate and the single-qp in our
case) present in tunnelling processes.

More delicate issue is the possibility to measure the precise values
of the excitation charges.  Indeed, as we saw before, one has to
consider low enough temperatures (blue curves in
figure~\ref{fig:Skewness}) in order to \emph{measure}, from the higher
plateau, the value of the charge associated with the $p$-agglomerate.
This can be explained by observing that for $V\to 0$, the current is
proportional to the linear conductances~(\ref{eq:conductance}),
consequently, from~(\ref{eq:Fksharing}) the cumulants are
\begin{equation}
\label{eq:Fkconduc}
F_k=\left(\frac{e^{*}}{e}\right)^{k-1}\f{G^{(1)} + (p)^{k-1}G^{(p)}}{G^{(1)}+G^{(p)}}\,.
\end{equation} 
As extensively discussed in~\cite{Ferraro08} one can verify that for temperatures low in comparison to the 
neutral modes bandwidth, $T\ll \omega_n$, the temperature power-law of the single-qp conductance is
$G^{(1)}\propto T^{2[g_c\nu/p^2+(1+1/p) g_n-1]}$ with positive exponent, while for the $p$-agglomerate one has 
$G^{(p)}\propto T^{2( g_c\nu-1)}$ with an exponent that changes sign for 
$g_c>1/\nu$.

Therefore for weak renormalization, $g_c<1/\nu$, decreasing temperature, the
$p$-agglomerate conductance strongly increase, and, at the same time,
the single-qp contribution becomes negligible. Then one does not need
extremely low temperatures in order to measure the true charge of $p$-agglomerate
from the value of the plateau.  This is well represented in
figure~\ref{fig:Skewness}(a) (green curve) where the skewness at low bias approaches the value
$\nu^2$. Note that this charge would have been more difficult to be
extracted from the Fano behaviour where the plateau is not so clearly
defined - see the same curve in figure~\ref{fig:Fano}(a) -. For strong
renormalization $g_c>1/\nu$, as in figure~\ref{fig:Skewness}(b), the
power-law exponents of the conductances $G^{(1)}$ and $G^{(p)}$ are
both positives. Then, the dominance of the $p$-agglomerate appears 
at much lower temperatures with respect to the previous case and the
plateau deviates from the $T=0$ value at temperatures lower than the weak
renormalization case.  In both case for temperatures not sufficiently low
the intermediate values of the higher plateau depend
on the weighted average of the two carrier contributions.

\section{Conclusion}
\label{conclusion}
We investigated the effects of propagating neutral modes in the tunnelling of quasiparticles
through a QPC for the Jain sequence of FQH. We observed that these modes
qualitatively modify the behaviour of the single-qp backscattering
current.  Even more interesting they determine the behaviour of the
total backscattering current given by the sum of the
single-qp and the $p$-agglomerate contribution.  
We demonstrated that the presence of neutral modes with finite
bandwidth creates a crossover between the two kinds of
quasiparticles involved in tunnelling.
To avoid non universal effects that may hide the possibility to 
extract information on this crossover we considered the current fluctuations.  
These quantities are indeed more robust than the current in order to identify unequivocally the entities which tunnel.
We first investigated the Fano factor that presents a plateaux structure related 
to the crossover between the single-qp and the
$p$-agglomerate tunnelling regime. However, the Fano is strongly affected by thermal
effects, therefore, we considered odd higher cumulants. We demonstrated that higher cumulants
are an important tool in order to establish the number of relevant
quasiparticle in the tunnelling. This is a strong motivation for
investigating these quantities, despite of the enormous
technical difficulties in their measurements. 

The cross-check of the physics described in this paper with some other
independent transport quantities such as the finite frequency noise~
\cite{Chamon95, Bena07, Zakka07} could open extremely interesting perspectives in the
comprehension of anomalous behaviours that emerge in noise
measurements. 
We are confident that analogous approaches could explain
some unclear experimental observation for also the case $\nu=5/2$.
\cite{Dolev08,Willet09}.
\ack We thank M. Heiblum, E. Fradkin, J. K. Jain, M. Dolev, H.-S. Sim and I. Safi for useful discussion. A. B. gratefully acknowledges the support of CNR-INFM by SEED project PLEASE001.

\appendix
\section{}
\label{App:BetaM}

This Appendix is devoted to the derivation of the coefficients
$\beta_m$ in~(\ref{beta1}). Let us start to discuss the 
monodromy condition introduced in section~\ref{Quasiparticle operators}.  
The generalization of the
internal statistical angle~(\ref{statistics}) for two different
agglomerates with $m$ and $m'$ is the so called  mutual statistical
angle $\Theta$ defined as~\cite{Ferraro09}
\begin{equation}
\label{A:mutualstatistics}
\Psi^{(m)}(x) \Psi^{(m')}(x')= \Psi^{(m')}(x') \Psi^{(m)}(x)e^{-i\  \Theta\  \textrm{sgn}(x-x')},
\end{equation}
the phase depends on the coefficients $(\alpha,\beta(k))$ and
$(\alpha',\beta'(k'))$ that define two different operators
in~(\ref{eq:quasiparticle_operator}) with (see
({\ref{alfa}) and (\ref{beta0}))
 \begin{equation}
 \alpha_{m}=\frac{m}{p}\,;\qquad
  \beta_{m}(k)=\sqrt{m^2\left(1+\frac{1}{p}\right)+2k},
\label{alfabeta}
\end{equation}
and similar expressions for $\alpha_{m'}$ and $\beta_{m'}(k')$.
By using the operator commutation rules one obtains from~(\ref{A:mutualstatistics})
\begin{equation}
\label{stat1}
\Theta [\alpha _{m} , \beta _{m}(k); \alpha _{m'},\beta _{m'}(k')] 
= \pi\left[\nu \alpha _{m} \alpha _{m'} - \beta _{m}(k) \beta _{m'}(k')\right].
\end{equation}

The monodromy condition imposes that the phase acquired by
any excitations $m$ in a loop around an electron must be a multiple of
$2 \pi$~\cite{Froehlich97,Ino98}.  We then need, first of all, to start
the discussion considering the electron operator, which corresponds to
$m=2p+1$. For this we will generalize the results presented
in~\cite{Ferraro09}, where the $\nu=2/5$ case was discussed.
In order to recover the electron operator of the original FL theory we consider 
$\beta_{2p+1}(k')$ in (\ref{alfabeta}) with
$k'=0$~\cite{Ferraro09}, we will check at the end the admissibility of this value.
Other electron operators in~(\ref{A:mutualstatistics}) with $k\not =0$ need to satisfy 
the mutual statistics constraint with the electron at $k'=0$. From~(\ref{stat1}) it is 
\begin{equation}
\Theta = \pi\left[\nu \alpha _{2p+1} \alpha _{2p+1} - \beta _{2p+1}(k) \beta _{2p+1}(0)\right]
= \pi r
\label{theta3}
\end{equation}
with $r\in\mathbb{Z}$.
Inserting (\ref{alfabeta}) in (\ref{theta3}) one has 
\begin{equation}
2 + \frac{1}{p}-\left(2+\frac{1}{p}\right)\sqrt{(p+1)(4p^3+8p^2+5p+1+2kp)}=r.
\label{Monodromy}
\end{equation}
To fulfil this equation one needs first of all to impose a constraint on the square root with
\begin{equation}
(p+1)(4p^3+8p^2+5p+1+2kp)=a^2.
\label{square_root}
\end{equation}
Hence, the monodromy condition~(\ref{Monodromy}) can be written as
\begin{equation}
\label{eqa}
\frac{2p+1}{p}(1-a)=r.
\end{equation}
One can see that to fulfil the two above equations one needs integer values of $a$.
The more general solution of (\ref{eqa}), is then written in terms of a number $u\in\mathbb{Z}$
\begin{equation}
a = 1+up, \qquad r=-u(2p+1).
\end{equation}
Replacing the above solution for $a$ in~(\ref{square_root}) one  has
\begin{equation}
pu^2 + 2u - (4p^3+12p^2+13p+6) = 2k (p+1).
\label{eqelectron}
\end{equation}
This represents a parabolic Diophantine equation for $u$ and $k$. In the following, we will restrict 
the analysis to the case with 
$p\leq6$. Here, the only possible solution, is given by
\begin{equation}
  u = 1+t(p+1),\qquad k = \frac{1}{2}(p+1)(t-2)(2+2p+pt),
\label{possible_k}
\end{equation}
where $t\in\mathbb{Z}$. For $p>6$  more complicated solutions are
present and will be not discussed in this paper. Note that the $t=2$ value
in~(\ref{possible_k}) gives $k=0$, in accordance with the initial choice $k'=0$
in~(\ref{theta3}).  Replacing the
expression~(\ref{possible_k}) for $k$ in~(\ref{alfabeta}), with $m=2p+1$ we obtain 
\begin{equation}
\alpha_{2p+1}=2+\frac{1}{p};\,\qquad
  \beta_{2p+1}(t)=\sqrt{p\left(p+1\right)}\left(t+\frac{1}{p}\right).
\label{general_electron}
\end{equation}

Now we generalize the analysis to the $m$-agglomerates.  The
monodromy condition~(\ref{stat1}) between an $m$-agglomerate described
by~(\ref{alfabeta}) and an electron, described
by~(\ref{general_electron}) is 
\begin{equation}
\frac{\Theta}{\pi}=\frac{m - \sqrt{2kp+m^2(p+1)}\sqrt{(p+1)}(pt+1)}{p}=v\,
\label{rel_stat_qp}
\end{equation}
where $v\in\mathbb{Z}$. 
Similarly to the previous case, one has first to impose a constraint on the square 
root~(\ref{rel_stat_qp})
\begin{equation}
\left[2kp+m^2(p+1)\right](p+1)=b^{2}.
\label{b}
\end{equation}
The monodromy condition is then written as
\begin{equation}
\frac{m-b(1+pt)}{p}=v.
\label{bb}
\end{equation}
One can see that to fulfil the two above equations one first need integer values of $b$.
The solution of~(\ref{bb}) is then
\begin{equation}
b = m-fp, \qquad v=f+(fp-m)t\qquad (f\in\mathbb{Z}).
\end{equation}
 Replacing the above solution for $b$ in~(\ref{b}) one  has
to solve the relation
\begin{equation}
f^{2}p-2fm - m^{2}(p+2)=2k(p+1),
\end{equation}
whose solution can be easily derived, again for $p\leq 6$,
\begin{equation}
f =-m-h(p+1),\qquad
k = h(p+1)(m+\frac{h}{2}p)
\label{akappa}
\end{equation}
with $k\in\mathbb{Z}$.

Note that for $m=2p+1$ and choosing $f=2-u$ one recover the
equation~(\ref{eqelectron}) for the electron. Through the
redefinitions $m = sp+d$ and $h = q-s$ $(q\in \mathrm{Z})$, we can
rewrite~(\ref{akappa}) as
\begin{equation}
\label{akappa1}
k=\frac{p\left(p+1\right)}{2}\left(q^{2}-s^{2}\right)+d\left(p+1\right)\left(q-s\right).
\end{equation}
This corresponds to the expression~(\ref{kappa1}) quoted in the main text.
Replacing~(\ref{akappa1}) in the initial expression~(\ref{alfabeta}) for $\beta_m(k)$ and using 
the notation $m = sp+d$ one has 
\begin{equation}
\beta_{m}(q)=\sqrt{p\left(p+1\right)}\left(q+\frac{d}{p}\right)
\label{general_coefficient}
\end{equation}
as defined in~(\ref{beta1}).  Note that for $s=2$, $d=1$ and
with the trivial identification $q=t$ we recover the expressions for the electron
in~(\ref{general_electron}).

\end{document}